\documentclass[10pt]{iopart}

\usepackage{graphicx}

\begin{document}

\title[Entanglement between two movable mirrors]
{Stationary entanglement between two movable mirrors in a
classically driven Fabry-Perot cavity}

\author{David Vitali
\footnote[3]{To whom correspondence should be addressed
(david.vitali@unicam.it)}, Stefano Mancini, and Paolo Tombesi }
\address{Dipartimento di Fisica,
Universit\`a di Camerino,
I-62032 Camerino, Italy
}

\begin{abstract}
We consider a Fabry-Perot cavity made by two moving mirrors and
driven by an intense classical laser field. We show that stationary
entanglement between two vibrational modes of the mirrors, with
effective mass of the order of micrograms, can be generated by means
of radiation pressure. The resulting entanglement is however quite
fragile with respect to temperature.
\end{abstract}





\section{Introduction}

Quantum entanglement is a physical phenomenon in which the quantum
states of two or more systems can only be described with reference
to each other. It is now intensively studied not just because of its
critical role in setting the boundary between classical and quantum
world, but also because it is an important physical resource that
allows performing communication and computation tasks with an
efficiency which is not achievable classically \cite{Nielsen}. In
particular, both from a conceptual and a practical point of view, it
is important to investigate under which conditions entanglement
between macroscopic objects, each containing a large number of the
constituents, can arise. Entanglement between two atomic ensembles
has been successfully demonstrated in Ref.~\cite{juuls01} by sending
pulses of coherent light through two atomic vapor cells. More
recently Ref.~\cite{Berkley} has shown spectroscopic evidence for
the creation of entangled macroscopic quantum states in two
current-biased Josephson-junction qubits coupled by a capacitor. The
interest has been also extended to micro- and nano-mechanical
oscillators, which have been shown to be highly controllable
\cite{LaHaye} and represent natural candidates for quantum limited
measurements and for testing decoherence theories \cite{marsh}.
Recent proposals suggested to entangle a nano-mechanical oscillator
with a Cooper-pair box \cite{Armour03}, arrays of nano-mechanical
oscillators \cite{eisert04}, two mirrors of an optical ring cavity
\cite{PRL02}, or two mirrors of two different cavities illuminated
with entangled light beams \cite{Peng03}. These two latter proposals
employed the optomechanical coupling provided by radiation pressure,
which has been demonstrated to provide a useful tool to manipulate
the quantum state of light
\cite{dorsel,bouwm,arci1,gigan,arci2,wien,kippen}.

\begin{figure}[htb]
\centerline{\includegraphics[width=0.75\textwidth]{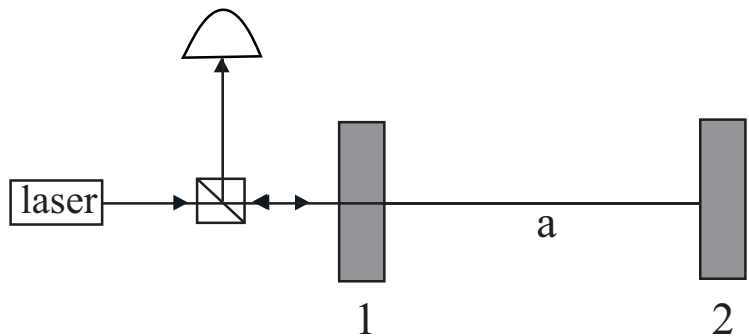}}
\caption{Schematic description of the system under study. The cavity
(with mode $a$) is driven by a laser, and the vibrating mirrors 1
and 2 are the ones we want to entangle.} \label{appa}
\end{figure}

Here we study the simplest scheme in which one can test the
entangling capabilities of radiation pressure, that is, a linear
Fabry-Perot cavity with two vibrating mirrors (see Fig.~\ref{appa}).
This system corresponds to a simplified version of the system of
Ref.~\cite{epl}, where a \emph{double-cavity} set-up formed by a
linear Fabry-Perot cavity and a ``folded'' ring cavity is
considered. Similarly to what has been done in Ref.~\cite{epl}, we
determine here the exact steady state of the system and show that if
the cavity is appropriately detuned, one can generate stationary
entanglement between macroscopic oscillators (effective mass $\sim
100$ ng). As it will be discussed below, the main advantages of the
present scheme with respect to that of Ref.~\cite{epl} are its
simplicity and the fact that steady-state entanglement is achievable
even with purely classical driving light, while Ref.~\cite{epl}
considered the limiting case of large mechanical frequencies where
entanglement can be generated only by injecting nonclassical
squeezed light into the two cavities.

The paper is organized as follows. In Section II we describe the
dynamics of the system in terms of quantum Langevin equations. In
Section III we solve the dynamics and derive the correlation matrix
of the steady state of the system. In Section IV we quantify the
mechanical entanglement in terms of the logarithmic negativity,
while in Section V we compare the present scheme with other recent
proposals for the generation of mechanical entanglement and discuss
how one can detect it. Section VI is for concluding remarks.

\section{The system}

We consider an optical Fabry-Perot cavity in which both mirrors can
move under the effect of the radiation pressure force (see
Fig.~\ref{appa}). The motion of each mirror is described by the
excitation of several degrees of freedom which have different
resonant frequencies. However, a single frequency mode can be
considered for each mirror when a bandpass filter in the detection
scheme is used \cite{Pinard} and mode-mode coupling is negligible.
Therefore we will consider a single mechanical mode for each mirror,
modeled as an harmonic oscillator with frequency $\Omega_k$ and
effective mass $M_k$, $k=1,2$, so that the mechanical Hamiltonian of
the mirrors is given by
\begin{equation}
H_m=\frac{P_1^2}{2M_1}+\frac{P_2^2}{2M_2}+
\frac{1}{2}M_1\Omega_1^2Q_1^2+\frac{1}{2}M_2\Omega_2^2P_2^2,
\end{equation}
with $[Q_k,P_j]=i\hbar \delta_{kj}$. In the adiabatic limit in which
the mirror frequencies are much smaller than the cavity free
spectral range $c/2L$ ($L$ is the cavity length in the absence of
the intracavity field) \cite{law}, one can focus on one cavity mode
only because photon scattering into other modes can be neglected,
and one has the following total Hamiltonian
\begin{equation} \label{ham0}
H=H_m+\hbar \omega_c a^{\dagger}a + \hbar \frac{\omega_c}{L}
a^{\dagger}a(Q_1-Q_2) + \imath\hbar E (e^{-\imath \omega_{L}
t}a^{\dag}-e^{\imath \omega_{L} t}a),
\end{equation}
where $a$ and $a^{\dag}$ ($[a,a^{\dag}]=1$) are the annihilation and
creation operators of the cavity mode with frequency $\omega_c$ and
decay rate $\kappa$, and the last two terms in Eq.~(\ref{ham0})
describe the driving laser with frequency $\omega_L$ and $E$ is
related to the input laser power $P$ by $|E|=\sqrt{2P \kappa/\hbar
\omega_L}$. In general the mirror potential is also determined by
the additional static Casimir term $V_{Cas}=-\hbar
\pi/\left[24(Q_2-Q_1+L)\right]$ \cite{law}, which however is
negligible for typical optical cavities with $L \sim 1$ cm and
mirrors with effective masses in the $\mu$g--ng range.

The full dynamics of the system is described by a set of nonlinear
Langevin equations, including the effects of vacuum radiation noise
and the quantum Brownian noise acting on the mirrors. In the
interaction picture with respect to $\hbar \omega_L a^{\dag}a$
\begin{eqnarray}
&&\dot{a}=-(\kappa+i\Delta_0)a-i\frac{\omega_c}{L} a(Q_1-Q_2)+E+\sqrt{2\kappa_1}a^{in}, \label{nonlinlang}\\
&& \dot{Q}_k = P_k/M_k, \;\;k=1,2,\\
&& \dot{P}_k=-M_k \Omega_k^2 Q_k-\gamma_k P_k+(-1)^k \hbar
\frac{\omega_c}{L} a^{\dagger} a +M_k W_k \;\;k=1,2,
\end{eqnarray}
where $\Delta_0=\omega_c-\omega_L$ and $\gamma_k$ is the mechanical
damping rate of mirror $k$. We have introduced the radiation input
noise $a^{in}(t)$, whose only nonzero correlation function is
\cite{gard}
\begin{equation}
 \langle a^{in}(t)a^{in,\dag}(t')\rangle =\delta (t-t'), \label{input}
\end{equation}
and the Hermitian Brownian noise operators $W_j(t)$, with zero mean
value and possessing the following correlation functions
\cite{Landau,GIOV01}
\begin{equation}\label{browncorre}
\left \langle W_{i}(t) W_{j}(t')\right \rangle =
\delta_{ij}\frac{\hbar \gamma_{j}}{M_j} \int_{-\infty}^{\infty}
  \frac{d\omega}{2\pi} e^{-i\omega(t-t')} \omega \left[\coth\left(\frac{\hbar \omega}{2k_BT}\right)+1\right],
\end{equation}
where $k_B$ is the Boltzmann constant and $T$ is the equilibrium
temperature, assumed to be equal for the two mirrors.

We are interested in the dynamics of the quantum fluctuations around
the steady state of the system. We can rewrite each Heisenberg
operator as a c-number steady state value plus an additional
fluctuation operator with zero mean value, $a = \alpha_s + \delta a
$, $Q_k = Q_k^s +  \delta Q_k $, $P_k = P_k^s +  \delta P_k$.
Inserting these expressions into the Langevin equations of
Eqs.~(\ref{nonlinlang}), these latter decouple into a set of
nonlinear algebraic equations for the steady state values and a set
of quantum Langevin equations for the fluctuation operators. The
steady state values are given by $P_k^s  =0$, ($k=1,2$),
$Q_k^s=(-1)^k(\hbar \omega_c/M_k\Omega_k^2L)|\alpha_s|^2$, $\alpha_s
= E/(\kappa+i \Delta)$, where the latter equation is in fact a
nonlinear equation determining the stationary intracavity field
amplitude $\alpha_s$, because the effective cavity detuning
$\Delta$, including radiation pressure effects, is given by
\begin{eqnarray}
&&\Delta = \Delta_0+\frac{\omega_c}{L}\left(Q_1^s-Q_2^s\right) \\
&&=\Delta_0-\hbar\left(\frac{\omega_c}{L}\right)^2|\alpha_s|^2\left(\frac{1}{M_1\Omega_1^2}+\frac{1}{M_2\Omega_2^2}\right).
\end{eqnarray}
The exact quantum Langevin equations for the fluctuations are
\begin{eqnarray}
&& \delta \dot{a}=-(\kappa+i\Delta)\delta a -i \frac{\omega_c}{L} \left(\alpha_s + \delta a\right)\left( \delta Q_1-\delta Q_2\right)
+\sqrt{2\kappa} a^{in}, \label{nlinlang1}\\
&&\delta \dot{Q_k}=\delta P_k/M_k \;\;k=1,2,\label{nlinlang2 }\\
&&\delta \dot{P_k}=-M_k \Omega_k^2 \delta Q_k-\gamma_k \delta
P_k+(-1)^k \hbar \frac{\omega_c}{L} \left(\alpha_s \delta
a^{\dagger}+ \alpha_s^* \delta a\right) \nonumber \\
&&+\delta a^{\dagger} \delta a + M_k W_k \;\;k=1,2.
\label{nlinlang3}
\end{eqnarray}
From a physical point of view the strong driving regime is the most
relevant one. In this regime, the intracavity amplitude is very
large, $|\alpha_s| \gg 1$, and, as shown by Eqs.~(\ref{nlinlang1})
and (\ref{nlinlang3}), one has a large effective optomechanical
coupling constant $\alpha_s \omega_c/L$ between the field quadrature
fluctuations and the oscillator. When $|\alpha_s| \gg 1$, one can
safely neglect the cavity field fluctuation operator $\delta a$ with
respect to $\alpha_s$ in Eqs.~(\ref{nlinlang1}) and
(\ref{nlinlang3}) and consider linearized Langevin equations. Notice
that this amounts to linearize only with respect to the cavity mode
and not with respect to the mechanical oscillator, whose operators
appear linearly in the dynamical equations from the beginning and
therefore are not approximated in the linearized treatment.

It is evident that the cavity mode is coupled only to the relative
motion of the two mirrors and it is therefore convenient to rewrite
the above equations in terms of the fluctuations of the relative and
center-of-mass coordinates, i.e.,
\begin{eqnarray}
\label{cenm}
\delta Q_{cm}&=&\frac{M_1}{M_T}\delta Q_1+\frac{M_2}{M_T}\delta Q_2,\hspace{0.5cm}\delta P_{cm}=\delta P_1+\delta P_2,\\
\delta Q_r&=&\delta Q_2-\delta Q_1,\hspace{0.5cm}\frac{\delta
P_r}{\mu}=\frac{\delta P_2}{M_2}-\frac{\delta P_1}{M_1},\label{relm}
\end{eqnarray}
where $M_T=M_1+M_2$ and $\mu=M_1M_2/M_T$ are the total and reduced
mass of the two oscillators. The linearized Langevin equations for
these coordinates are
\begin{eqnarray}
\delta \dot{a}&=&-(\kappa+i\Delta)\delta a +i \frac{\omega_c}{L} \alpha_s \delta Q_r +\sqrt{2\kappa} a^{in},\\
\delta \dot{Q}_r &=& \delta P_r/\mu,\\
\delta \dot{P}_r &=& -\mu \Omega_r^2 \delta Q_r-\gamma_r \delta
P_r-\mu \left(\Omega_2^2-\Omega_1^2\right)\delta Q_{cm} \nonumber \\
&&-\frac{\mu}{M_T}\left(\gamma_2-\gamma_1\right)
\delta P_{cm} +\hbar \frac{\omega_c}{L}\left(\alpha_s^* \delta a+\alpha_s \delta a^{\dagger}\right) +\mu W_r ,\\
\delta \dot{Q}_{cm}&=& \delta P_{cm}/M_T,\\
\delta \dot{P}_{cm}&=&-M_T \Omega_{cm}^2 \delta Q_{cm}-\gamma_{cm}
\delta P_{cm}\nonumber \\
&&-\mu \left(\Omega_2^2-\Omega_1^2\right)\delta Q_r
-\left(\gamma_2-\gamma_1\right)\delta P_r +M_T W_{cm},
\end{eqnarray}
where we have defined the center-of-mass frequency
$\Omega_{cm}^2=\left(M_1\Omega_1^2+M_2\Omega_2^2\right)/M_T$,
damping rate $\gamma_{cm}=\left(M_1\gamma_1+M_2\gamma_2\right)/M_T$,
and Brownian noise $W_{cm}=\left(M_1W_1+M_2W_2\right)/M_T$, and also
the relative motion frequency
$\Omega_{r}^2=\left(M_2\Omega_1^2+M_1\Omega_2^2\right)/M_T$, damping
rate $\gamma_{r}=\left(M_2\gamma_1+M_1\gamma_2\right)/M_T$ and
Brownian noise $W_r=W_2-W_1$. Thanks to these definitions, the
center-of-mass and relative motion Brownian noise possess
correlation functions analogous to those of Eq.~(\ref{browncorre}),
with the corresponding damping rate and mass. The two noises are
however correlated in general, because
\begin{eqnarray}
&& \left \langle W_{cm}(t) W_{r}(t')\right \rangle = \left \langle
W_{r}(t) W_{cm}(t')\right \rangle \nonumber \\
&& =\frac{\hbar (\gamma_{2}-\gamma_{1})}{M_T}\int_{-\infty}^{\infty}
  \frac{d\omega}{2\pi} e^{-i\omega(t-t')} \omega \left[\coth\left(\frac{\hbar \omega}{2k_BT}\right)+1\right].
\label{browncorre4}
\end{eqnarray} The above equations show that,
even though the cavity mode directly interacts only with the
relative motion, the three modes are all coupled because of the
center-of-mass--relative-motion coupling, which is present whenever
$\Omega_1\neq \Omega_2$ or $\gamma_1\neq \gamma_2$.

\subsection{Equal frequencies and damping rates}

The dynamics considerably simplify when $\Omega_1= \Omega_2=\Omega$
and $\gamma_1= \gamma_2=\gamma$. In fact, in such a case
$\Omega_{cm}=\Omega_r=\Omega$ and $\gamma_{cm}=\gamma_r=\gamma$ and
the center-of-mass motion fully decouples from the cavity mode and
the relative motion, even if the masses are different. The
center-of-mass becomes an isolated quantum oscillator with mass
$M_T$ and subject to quantum Brownian noise, i.e.,
\begin{eqnarray}
&& \delta \dot{Q}_{cm} = \delta P_{cm}/M_T,\\
&& \delta \dot{P}_{cm}=-M_T \Omega^2 \delta Q_{cm}-\gamma \delta
P_{cm} +M_T W_{cm},
\end{eqnarray}
while the relative position of the two mirrors and the linearized
fluctuations of the cavity mode form a system of two interacting
modes described by the following linear Langevin equations
\begin{eqnarray}
&& \delta \dot{Q}_r = \delta P_r/\mu,\label{linle} \\
&& \delta \dot{P}_r=-\mu \Omega^2 \delta Q_r-\gamma \delta P_r
+ \frac{\sqrt{2}\hbar\omega_c \alpha_s}{L} X +\mu W_r ,\\
&& \dot{X}=-\kappa X+\Delta Y +\sqrt{2\kappa} X^{in},  \\
&& \dot{Y}=-\kappa Y-\Delta X +\frac{\sqrt{2}\omega_c \alpha_s}{L}
\delta Q_r +\sqrt{2\kappa} Y^{in},
\end{eqnarray}
where we have chosen the phase reference of the cavity field so that
$\alpha_s$ is real, we have defined the cavity field quadratures $
X\equiv(\delta a+\delta a^{\dag})/\sqrt{2}$ and $ Y\equiv(\delta
a-\delta a^{\dag})/i\sqrt{2}$, and the corresponding Hermitian input
noise operators $X^{in}\equiv(a^{in}+a^{in,\dag})/\sqrt{2}$ and
$Y^{in}\equiv(a^{in}-a^{in,\dag})/i\sqrt{2}$. Notice that
Eqs.~(\ref{linle}) coincide with the linearized equations of a
Fabry-Perot cavity with only \emph{one} movable mirror with mass
$\mu$.

It is convenient to switch to dimensionless dynamical variables for
the mechanical oscillators. If we define
\begin{eqnarray}
&& \delta Q_{k}=\sqrt{\frac{\hbar}{M_k \Omega}}q_k,\hspace{0.5cm}\delta P_{k}=\sqrt{\hbar M_k \Omega}p_k,\; k=1,2, \label{adim}\\
&& \delta Q_{cm}=\sqrt{\frac{\hbar}{M_T \Omega}}q_{cm},\hspace{0.5cm}\delta P_{cm}=\sqrt{\hbar M_T \Omega}p_{cm}, \\
&& \delta Q_{r}=\sqrt{\frac{\hbar}{\mu
\Omega}}q_{r},\hspace{0.5cm}\delta P_{r}=\sqrt{\hbar \mu
\Omega}p_{r},
\end{eqnarray}
such that $[q_j,p_k]=i\delta_{jk}$, either for $j,k=1,2$ and for
$j,k=r,cm$, definitions (\ref{cenm})-(\ref{relm}) become
\begin{eqnarray}
q_{cm}&=&r_1 q_1+r_2 q_2,\hspace{0.5cm}p_{cm}=r_1 p_1+r_2 p_2, \label{ceadim}\\
q_{r}&=&r_1 q_2-r_2 q_1,\hspace{0.5cm}p_{r}=r_1 p_2-r_2 p_1,
\end{eqnarray}
where $r_k=\sqrt{M_k/M_T}$, $k=1,2$. The quantum Langevin equations
become in terms of these dimensionless continuous variables
\begin{eqnarray}
&& \dot{q}_{cm} = \Omega p_{cm},\label{linleadim}\\
&& \dot{p}_{cm}=-\Omega q_{cm}-\gamma p_{cm} + \xi_{cm}, \\
&& \dot{q}_r = \Omega p_{r},\\
&& \dot{p}_r= \Omega q_r-\gamma p_r + G X + \xi_{r},\\
&& \dot{X}=-\kappa X+\Delta Y +\sqrt{2\kappa} X^{in},  \\
&& \dot{Y}=-\kappa Y-\Delta X +G q_r +\sqrt{2\kappa} Y^{in},
\end{eqnarray}
where we have defined the effective optomechanical coupling constant
\begin{equation}
G=\sqrt{\frac{2\hbar}{\mu \Omega}}\frac{\alpha_s
\omega_c}{L}=\frac{2\omega_c}{L}\sqrt{\frac{P \kappa}{\mu \Omega
\omega_L \left(\kappa^2+\Delta^2\right)}}, \label{optoc}
\end{equation}
which, being proportional to the square root of the input power, can
be made quite large, and the zero-mean scaled Brownian noise
operators $\xi_r(t)$ and $\xi_{cm}(t)$, with correlation functions
\begin{equation}\label{browncorre5}
\left \langle \xi_j(t) \xi_k(t')\right \rangle
=\delta_{jk}\frac{\gamma}{\Omega}\int_{-\infty}^{\infty}
  \frac{d\omega}{2\pi} e^{-i\omega(t-t')} \omega \left[\coth\left(\frac{\hbar
  \omega}{2k_BT}\right)+1\right]\;,
\end{equation}
where $j,k=$r,cm.

\section{Stationary correlation matrix of the two mirrors}

When the three-mode system is stable, it reaches a unique steady
state, independently from the initial condition. Since the quantum
noises $\xi_{cm}$, $\xi_{r}$, $X^{in}$ and $Y^{in}$ are zero-mean
quantum Gaussian noises and the dynamics is linearized, the quantum
steady state for the fluctuations is a zero-mean Gaussian state,
fully characterized by its $6 \times 6 $ correlation matrix (CM) $
V^{(6)}_{ij}=\langle u_i(\infty)u_j(\infty)+
u_j(\infty)u_i(\infty)\rangle/2$, where $u^{T}(\infty) =
(q_1(\infty), p_1(\infty), q_2(\infty), p_2(\infty), X(\infty),
Y(\infty))$ is the vector of continuous variables (CV) fluctuation
operators at the steady state ($t \to \infty$). We are interested in
the stationary reduced state of the two mirrors, which is obtained
by tracing out the cavity mode. This state is obviously still
Gaussian and fully characterized by the $4 \times 4$ matrix $V_{ij}$
formed by the first four rows and columns of $V^{(6)}_{ij}$. The
general form of $V$ is quite simple. First of all it is
$V_{12}=V_{34}=0$. In fact, since $p_j=\dot{q}_j/\Omega$, $j=1,2$,
it is
\begin{equation}
V_{12}=\frac{\langle q_1(\infty)p_1(\infty)+
p_1(\infty)q_1(\infty)\rangle}{2} = \frac{1}{2\Omega}\lim_{t\to
\infty}\frac{d}{dt}\langle q_1^2(t)\rangle =0,
\end{equation}
and the same happens for $V_{34}$. Moreover, thanks to the
decoupling between center-of-mass and relative motion it is
$V_{14}=V_{23}=0$, because
\begin{eqnarray}
&& V_{14}=\frac{\langle q_1(\infty)p_2(\infty)+
p_2(\infty)q_1(\infty)\rangle}{2} \nonumber \\
&&=\frac{r_1 r_2}{2}\left[\langle q_{cm}(\infty)p_{cm}(\infty)+
p_{cm}(\infty)q_{cm}(\infty)\rangle
\right. \nonumber \\
&&-\left. \langle q_{r}(\infty)p_{r}(\infty)+
p_{r}(\infty)q_{r}(\infty)\rangle\right]\nonumber \\
&&= \frac{r_1 r_2}{2\Omega }\lim_{t\to
\infty}\frac{d}{dt}\left[\langle q_{cm}^2(t)\rangle -\langle
q_{r}^2(t)\rangle\right]=0,
\end{eqnarray}
and the same happens for $V_{23}$. The final form of $V$ is
\begin{equation}\label{cmv}
  V=\left(\begin{array}{cccc}
    V_{11}& 0 & V_{13} & 0 \\
     0 & V_{22} & 0 & V_{24} \\
    V_{13} & 0 & V_{33} & 0 \\
    0 & V_{24} & 0 & V_{44}
  \end{array}\right),
\end{equation}
where
\begin{eqnarray}
V_{11}&=&r_1^2\langle q_{cm}^2\rangle_{st}+ r_2^2\langle q_{r}^2\rangle_{st},\hspace{0.5cm}
V_{22}=r_1^2\langle p_{cm}^2\rangle_{st}+ r_2^2\langle p_{r}^2\rangle_{st}, \label{V11V22}\\
V_{33}&=&r_2^2\langle q_{cm}^2\rangle_{st}+ r_1^2\langle
q_{r}^2\rangle_{st},\hspace{0.5cm}
V_{44}=r_2^2\langle p_{cm}^2\rangle_{st}+ r_1^2\langle p_{r}^2\rangle_{st}, \label{V33V44}\\
V_{13}&=&r_1 r_2\left(\langle q_{cm}^2\rangle_{st}-\langle
q_{r}^2\rangle_{st}\right),\hspace{0.5cm} V_{24}=r_1
r_2\left(\langle p_{cm}^2\rangle_{st}-\langle
p_{r}^2\rangle_{st}\right), \label{V13V24}
\end{eqnarray}
that is, it depends upon the mass ratios $r_k$ and the four
stationary variances $\langle u^2\rangle_{st}\equiv \lim_{t \to
\infty} \langle u^2(t)\rangle$, $u=q_{cm}, p_{cm},q_r,p_r$.

\subsection{Calculation of the stationary variances}

The center-of-mass and relative motion stationary variances can be
obtained by solving Eqs.~(\ref{linleadim}) and considering the limit
$t \to \infty$. Defining the six-dimensional vector of variables
$v^{T}(t) =(q_{cm}(t), p_{cm}(t),q_{r}(t), p_{r}(t),X(t),Y(t))$, the
vector of noises $n^{T}(t) =(0, \xi_{cm}(t),0,
\xi_r(t),\sqrt{2\kappa}X^{in}(t), \sqrt{2\kappa}Y^{in}(t))$ and the
matrix
\begin{equation}\label{dynmat}
  A=\left(\begin{array}{cccccc}
    0 & \Omega & 0 & 0 & 0 & 0\\
     -\Omega & -\gamma & 0 & 0 & 0 & 0\\
     0 & 0 & 0 & \Omega & 0 & 0\\
     0 & 0 & -\Omega & -\gamma & G & 0 \\
    0 & 0 & 0 & 0 & -\kappa & \Delta \\
    0 & 0 & G & 0 & -\Delta & -\kappa
  \end{array}\right),
\end{equation}
Eqs.~(\ref{linleadim}) can be rewritten in compact form as
$\dot{v}(t)=A v(t)+n(t)$, whose solution is
\begin{equation} \label{dynsol}
v(t)=M(t) v(0)+\int_0^t ds M(s) n(t-s),
\end{equation}
where $M(t)=\exp\{A t\}$. The system is stable and reaches its
steady state when all the eigenvalues of $A$ have negative real
parts so that $M(\infty)=0$. The stability conditions can be derived
by applying the Routh-Hurwitz criterion \cite{grad}, yielding the
following two nontrivial conditions on the system parameters,
\begin{eqnarray}
&& s_1=2 \gamma \kappa \left[\Delta^4 +\Delta^2(\gamma^2+2\gamma \kappa +2 \kappa^2-2\Omega^2)\right. \label{stab}\\
&&\left.+(\gamma \kappa +\kappa^2+\Omega^2)^2\right]+ \Omega G^2
\Delta
(\gamma+2\kappa)^2 > 0 , \nonumber \\
&& s_2=\Omega\left(\Delta^2+\kappa^2\right)-G^2 \Delta > 0,
\end{eqnarray}
which will be considered to be satisfied from now on. If we consider
the variables $v_j(t)$, we can construct the stationary correlation
matrix
\begin{equation}
C_{ij}=\frac{\langle
v_i(\infty)v_j(\infty)+v_j(\infty)v_i(\infty)\rangle}{2},
\end{equation}
which is the quantity of interest because $C_{11}=\langle
q_{cm}^2\rangle_{st}$, $C_{22}=\langle p_{cm}^2\rangle_{st}$,
$C_{33}=\langle q_{r}^2\rangle_{st}$, and $C_{44}=\langle
p_{r}^2\rangle_{st}$. When the system is stable, using
Eq.~(\ref{dynsol}) one gets
\begin{equation} \label{cm2}
C_{ij}=\sum_{k,l}\int_0^{\infty} ds \int_0^{\infty}ds' M_{ik}(s)
M_{jl}(s')\Phi_{kl}(s-s'),
\end{equation}
where $\Phi_{kl}(s-s')=\left(\langle n_k(s)n_l(s')+
n_l(s')n_k(s)\rangle\right)/2$ is the matrix of the stationary noise
correlation functions. Due to Eq.~(\ref{browncorre5}), the mirror
Brownian noises are not delta-correlated and therefore do not
describe in general a Markovian process. However, as we shall see,
mechanical entanglement is achievable only using oscillators with a
very good mechanical quality factor $\mathcal{Q}=\Omega/\gamma$. In
this weak damping limit, $\gamma \to 0$, the quantum Brownian noises
$\xi_r(t)$ and $\xi_{cm}(t)$ become delta-correlated,
\cite{benguria}
\begin{equation}\label{browncorre6}
\left \langle \xi_j(t) \xi_k(t')+\xi_k(t') \xi_j(t)\right \rangle/2
\simeq \delta_{jk} \gamma \left(2\bar{n}+1\right) \delta(t-t'),
\end{equation}
where $\bar{n}=\left(\exp\{\hbar \Omega/k_BT\}-1\right)^{-1}$, is
the mean thermal excitation number, and one recovers a Markovian
process. Using the definitions of $X_{in}$ and $Y_{in}$ and
Eq.~(\ref{input}), we finally get $\Phi_{kl}(s-s')= D_{kl}
\delta(s-s')$, where
\begin{equation}\label{diago}
  D=\left(\begin{array}{cccc}
    0 & 0 & 0 & 0 \\
     0 & \gamma(2\bar{n}+1) & 0 & 0 \\
    0 & 0 & \kappa & 0 \\
    0 & 0 & 0 & \kappa
  \end{array}\right).
\end{equation}
As a consequence, Eq.~(\ref{cm2}) becomes
\begin{equation} \label{cm3}
C =\int_0^{\infty} ds  M(s)D M(s)^{T},
\end{equation}
which, when the stability conditions are satisfied so that
$M(\infty)=0$, is equivalent to the following equation for the CM,
\begin{equation} \label{lyap}
AC+CA^{T}=-D.
\end{equation}
Eq.~(\ref{lyap}) is a linear equation for $C$ and it can be
straightforwardly solved. The center-of-mass is decoupled from the
other two modes and Eq.~(\ref{lyap}) trivially gives
\begin{equation} \label{cmst}
C_{11}=C_{22}=\langle q_{cm}^2\rangle_{st}=\langle
p_{cm}^2\rangle_{st}=\frac{1}{2}+\bar{n}.
\end{equation}
The relative motion is instead coupled with the cavity mode and
consequently the final expression of the stationary variances are
much more involved. One has
\begin{eqnarray}
C_{33}&=&\langle q_{r}^2\rangle_{st}=\frac{1}{2}+b_q+d_q \bar{n}, \label{qrel} \\
C_{44}&=&\langle p_{r}^2\rangle_{st}=\frac{1}{2}+b_p+d_p \bar{n},
\label{prel}
\end{eqnarray}
where
\begin{eqnarray}
b_p & =& [s_1]^{-1}G^2 \kappa \left\{\Delta^2(\gamma+\kappa)+\kappa
(\gamma \kappa +\kappa^2+\Omega^2)-\Delta\Omega (\gamma+2\kappa)
\right\}, \label{bip}\\
d_p & =& 1-[s_1]^{-1} 2 G^2 \kappa \Omega \Delta (\gamma+2\kappa), \label{dip}\\
b_q & =& [2s_1s_2]^{-1}G^2 \left\{ 2 \kappa \left( \Delta^2 +
\kappa^2 \right) \left\{ \left[ \Delta^2 + {\left( \gamma + \kappa
\right) }^2 \right]\left(\kappa\Omega+\gamma
\Delta\right)\right. \right.\label{biq}\\
&+ & \left. \left.
\Omega^2\left(\gamma+\kappa\right)\left(\Omega-2\Delta\right)\right\}
\Delta G^2\Omega^2\left(\gamma+2\kappa\right)\left[\Delta \gamma-\kappa \left(\Omega-2\Delta\right)\right]\right\}, \nonumber \\
d_q & =& 1+[s_1s_2]^{-1}\Delta G^2\left[s_1-2\gamma\kappa
\Omega^2\left(\Omega^2+2\gamma \kappa +4 \kappa^2\right)\right.
\nonumber \\
&-& \left. 4\kappa^2 \Omega^2\left(\Delta^2+\kappa^2\right)\right].
\label{diq}
\end{eqnarray}

\section{Conditions for stationary entanglement}

Simon's separability PPT (positive partial transpose) criterion is
necessary and sufficient for bipartite Gaussian CV states
\cite{simon}. It assumes a particularly simple form for the CM of
the two mirrors of Eq.~(\ref{cmv}). In fact, after some algebra, one
gets the following necessary and sufficient condition for the
presence of mechanical entanglement between the two mirrors in the
stationary state,
\begin{eqnarray}
  &&\left[\langle q_{r}^2\rangle_{st}\langle p_{cm}^2\rangle_{st}-\frac{1}{4}\right]
  \left[\langle p_{r}^2\rangle_{st}\langle q_{cm}^2\rangle_{st}-\frac{1}{4}\right]<
  \left(1-\frac{1}{\eta}\right) \nonumber \\
  &&\times \left[\langle q_{cm}^2\rangle_{st}\langle p_{cm}^2\rangle_{st}-\frac{1}{4}\right]
  \left[\langle q_{r}^2\rangle_{st}\langle
  p_{r}^2\rangle_{st}-\frac{1}{4}\right], \label{simoncrit}
\end{eqnarray}
where we have defined $\eta=4r_1^2 r_2^2=4\mu/M_T$. For very
different masses $\eta \to 0$ and the right hand side of
Eq.~(\ref{simoncrit}) tends to $-\infty$, i.e., the criterion is
never satisfied and the mirrors are never entangled. It is evident
therefore that stationary entanglement is better achieved for
\emph{equal mirrors}, i.e., $\eta =1$, when the right hand side of
Eq.~(\ref{simoncrit}) is equal to zero and the necessary and
sufficient entanglement condition becomes equivalent to a
``product'' of sufficient criteria analogous to those derived in
\cite{PRL02,noi}, that is, $\langle q_{r}^2\rangle_{st}\langle
p_{cm}^2\rangle_{st} < 1/4$ or $\langle p_{r}^2\rangle_{st}\langle
q_{cm}^2\rangle_{st} < 1/4$. Since the center-of-mass of the two
mirror is unaffected by the optomechanical coupling (see
Eq.~(\ref{cmst})), this means that the two mirror vibrational modes
are entangled if and only if their relative motion is sufficiently
squeezed, i.e.,
\begin{equation}\label{squeecond}
\langle q_{r}^2\rangle_{st} \;\; {\rm or}\;\; \langle
p_{r}^2\rangle_{st} < \frac{1}{2(1+2\bar{n})}.
\end{equation}
This equation provides an intuitive picture of how the entanglement
between the two mirrors is generated by the radiation pressure of
the light bouncing between them. If the cavity is strongly driven,
the radiation pressure coupling becomes very large and the
fluctuations of the mirror relative motion can be significantly
squeezed. If such a squeezing is large enough to overcome even the
thermal noise acting on the center-of-mass, Eq.~(\ref{squeecond})
guarantees that the two mirrors are entangled. Eq.~(\ref{squeecond})
also points out the main limit of the proposed scheme: the mirrors
center-of-mass is not affected by radiation pressure and cannot be
squeezed. This suggests that the generated entanglement is not
robust against temperature because satisfying Eq.~(\ref{squeecond})
becomes prohibitive at large $\bar{n}$.

One can quantify the stationary mechanical entanglement by
considering the logarithmic negativity $E_{\mathcal{N}}$
\cite{werner}, which in the CV case $E_{\mathcal{N}}$ can be defined
as \cite{Salerno1}
\begin{equation}
E_{\mathcal{N}}=\max [0,-\ln 2\nu ^{-}],  \label{logneg}
\end{equation}
where $\nu^{-}$ is given by
\begin{equation}
\nu ^{-}\equiv 2^{-1/2}\left[ \Sigma (V)- \left( \Sigma
(V)^{2}-4\det V\right)^{1/2}\right]^{1/2}, \label{Sympl_eigenv}
\end{equation}
with $\Sigma (V)\equiv \det N_1+\det N_2-2\det N_{12}$ and we have
used the $2\times2$ block form of the CM
\begin{equation}
V\equiv \left(
\begin{array}{cc}
N_1 & N_{12} \\
N_{12}^{T} & N_2
\end{array}
\right) .  \label{blocks}
\end{equation}
Therefore, a Gaussian state is entangled if and only if $ \nu
^{-}<1/2$, which is equivalent to Simon's necessary and sufficient
entanglement criterion for Gaussian states \cite{simon} of
Eq.~(\ref{simoncrit}), and which can be written as $4\det V <
\Sigma(V) -1/4$. In the case of the stationary matrix $V$ of
Eq.~(\ref{cmv}), one has
\begin{eqnarray}
&& \det V = \langle q_{r}^2\rangle_{st} \langle p_{r}^2\rangle_{st} \langle q_{cm}^2\rangle_{st} \langle p_{cm}^2\rangle_{st} \\
&& \Sigma(V)=(1-\eta)\left[\langle q_{r}^2\rangle_{st} \langle p_{r}^2\rangle_{st}+ \langle q_{cm}^2\rangle_{st} \langle p_{cm}^2\rangle_{st}\right] \nonumber \\
&& +\eta \left[\langle q_{r}^2\rangle_{st} \langle
p_{cm}^2\rangle_{st}+ \langle q_{cm}^2\rangle_{st} \langle
p_{r}^2\rangle_{st}\right].
\end{eqnarray}
Therefore, in the most convenient condition for entanglement, i.e.,
identical mirrors $\Leftrightarrow \eta=1$, one has
$\Sigma(V)=\langle q_{r}^2\rangle_{st} \langle p_{cm}^2\rangle_{st}
+ \langle q_{cm}^2\rangle_{st} \langle p_{r}^2\rangle_{st}$,
yielding
\begin{equation}
\nu^{-}=\min\left\{\sqrt{\langle q_{r}^2\rangle_{st} \langle
p_{cm}^2\rangle_{st}}, \sqrt{\langle q_{cm}^2\rangle_{st} \langle
p_{r}^2\rangle_{st}}\right\},
\end{equation}
so that in this case of equal masses, the logarithmic negativity
assumes the particularly simple form
\begin{equation}
E_{\mathcal{N}}=\max \left\{0,-\ln \left[2\sqrt{\langle
q_{r}^2\rangle_{st} \langle p_{cm}^2\rangle_{st}}\right], -\ln
\left[2\sqrt{\langle q_{cm}^2\rangle_{st} \langle
p_{r}^2\rangle_{st}}\right]\right\}.  \label{logneg2}
\end{equation}
Using Eqs.~(\ref{cmst})-(\ref{prel}), and (\ref{logneg2}) one has
stationary entanglement if one of the two following conditions is
satisfied
\begin{eqnarray}
&& b_q+d_q \bar{n} <  -\frac{\bar{n}}{2\bar{n}+1},  \label{qrelcond} \\
&& b_p+d_p \bar{n} <  -\frac{\bar{n}}{2\bar{n}+1}, \label{prelcond}
\end{eqnarray}
which, as expected, are better satisfied in the zero temperature
limit, $\bar{n} \to 0$, since $d_q,d_p\geq 0$ whenever the stability
conditions are satisfied (otherwise one could have negative
variances at large enough temperatures).

\begin{figure}[htb]
\centerline{\includegraphics[width=0.75\textwidth]{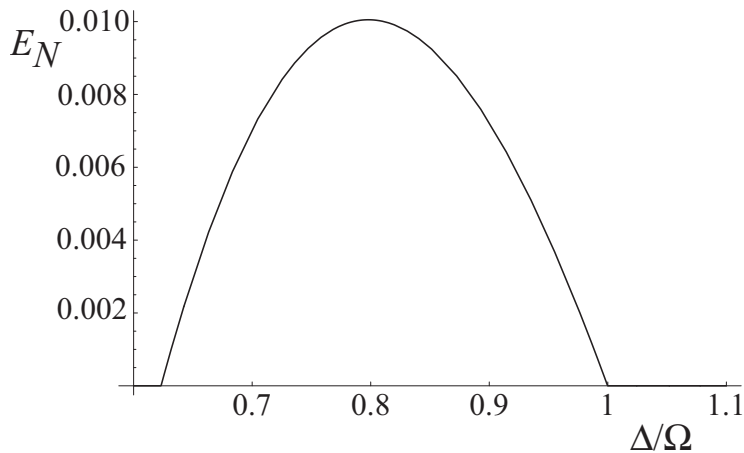}}
\caption{Logarithmic negativity $E_{\mathcal{N}}$ as a function of
the normalized detuning $\Delta/\Omega$ in the case of an optical
cavity of length $L=1$ cm, finesse $\mathcal{F}=1.9 \times 10^5$,
driven by a laser with wavelength $1064$ nm and power $P=50$ mW. The
identical movable mirrors have a frequency $\Omega/2\pi=10$ MHz,
damping rate $\gamma=3 \times 10^5$ s$^{-1}$, mass $m=100$ ng and
their temperature is $T=0$.} \label{entsi}
\end{figure}

\begin{figure}[htb]
\centerline{\includegraphics[width=0.75\textwidth]{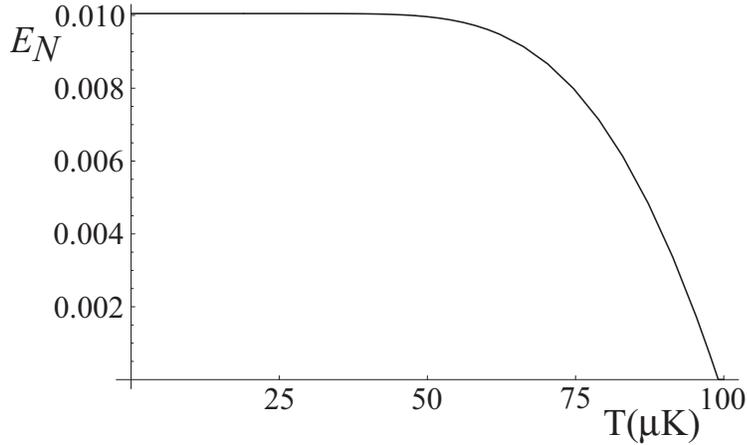}}
\caption{Logarithmic negativity $E_{\mathcal{N}}$ evaluated at
$\Delta=\Delta_{opt} \simeq 0.8 \Omega $ (see
Eq.~\protect\ref{dopt})) as a function of temperature. The other
parameter values are the same as in Fig.~\protect\ref{entsi}.
}\label{entno}
\end{figure}

These two equations lead us to the main result of the paper, i.e.,
it is possible to realize an entangled stationary state of two
macroscopic movable mirrors of a classically driven Fabry-Perot
cavity. However, such a stationary mechanical entanglement turns out
to be fragile with respect to temperature, as it can be easily
grasped from Eqs.~(\ref{qrelcond})-(\ref{prelcond}). This is
illustrated in Figs.~\ref{entsi}-\ref{entno}, where we have
considered a parameter region very close to that of recently
performed experiments employing optical Fabry-Perot cavities with at
least one micromechanical mirror \cite{bouwm,arci1,gigan,arci2}.
Figs.~\ref{entsi}-\ref{entno} refer to the case of an optical cavity
of length $L=1$ cm, finesse $\mathcal{F}=1.9 \times 10^5$, so that
$\kappa \simeq 5 \times 10^5$ s$^{-1}$, driven by a laser with
wavelength $1064$ nm and power $P=50$ mW. The two identical
mechanical oscillators have angular frequency $\Omega/2\pi=10$ MHz,
damping rate $\gamma=3 \times 10^5$ s$^{-1}$, and mass $m=100$ ng.
Fig.~\ref{entsi} refers to the zero temperature limit and shows that
stationary entanglement is present only within a small interval of
values of $\Delta$ around $\Delta \simeq \Delta_{opt}$ where
\begin{equation}\label{dopt}
\Delta_{opt}= \Omega \frac{\gamma+2\kappa}{2\gamma+2\kappa}.
\end{equation}
This value is essentially the optimal value for the detuning for
achieving entanglement. This can be understood from the expression
of $\langle p_{r}^2\rangle_{st}$. In fact, at zero temperature
entanglement is obtained when $b_p <0$ (see Eq.~(\ref{prelcond})),
which is satisfied when the numerator of Eq.~(\ref{bip}) is
negative, since $s_1 >0$ due to stability. This condition is
obtained by considering the minimum of the second order polynomial
in $\Delta$ in the numerator and by imposing that it is negative.
The minimum value of this polynomial is obtained just at $\Delta
=\Delta_{opt}$ and it is negative when
\begin{equation}\label{sufcond}
\gamma \Omega > 2\kappa (\gamma+\kappa).
\end{equation}
Therefore $\Delta = \Delta_{opt}$ and Eq.~(\ref{sufcond}) are
sufficient conditions for achieving entanglement at zero
temperature. This parameter regime is the optimal for entanglement
because when $\Delta \simeq \Delta_{opt} \simeq \Omega$, $s_1$ is
also close to its minimum value, implying therefore a large negative
value of $b_p$ (see Eq.~(\ref{bip})) and also a value of $d_p$ very
close to zero (see Eq.~(\ref{dip})), which means an improved
robustness of entanglement with respect to temperature. In
Fig.~\ref{entno} we study the resistance to thermal effects by
plotting $E_{\mathcal{N}}$ evaluated at the optimal detuning, i.e.,
corresponding to the maximum of Fig.~\ref{entsi}, versus
temperature. We see that this entanglement vanishes for $T > 100$
$\mu$K. This behavior is valid in general, even in parameter regions
different from that of Figs.~\ref{entsi}, \ref{entno}: whenever one
finds a regime with a nonzero stationary entanglement, this
entanglement quickly tends to zero for increasing temperatures. As
discussed above (see below Eq.~(\ref{squeecond})) this is due to the
fact that in this simple Fabry-Perot cavity system, the mirror
center-of-mass is unaffected by the radiation pressure of the cavity
mode and remains at thermal equilibrium. One could achieve a larger
and more robust entanglement by adopting the double-cavity setup
considered in \cite{epl}, where the optical mode of the second,
``folded'' cavity couples just to the center-of-mass of the mirrors
of interest, which is then also squeezed, independently from the
relative motion. In this latter scheme therefore robustness against
temperature is achieved at the price of a much more involved
experimental setup.

Eq.~(\ref{logneg2}) shows that mechanical entanglement at zero
temperature could be realized as well when $\langle
q_{r}^2\rangle_{st}< 1/2$. However, it is possible to see through
numerical calculations that this condition is much more difficult to
realize with respect to $\langle p_{r}^2\rangle_{st}< 1/2$. This
fact is not easily seen from the analytical expressions of $b_q$ and
$d_q$ (Eqs.~(\ref{biq})-(\ref{diq})), which are more difficult to
analyze with respect to those of $b_p$ and $d_p$
(Eqs.~(\ref{bip})-(\ref{dip})).

\section{Comparison with other proposals and experimental detection
of the entanglement}

It is interesting to compare the present proposal with other recent
schemes for entangling two micro-mechanical mirrors, especially with
Refs.~\cite{PRL02,Peng03,epl,entswap}, which are all based on the
optomechanical coupling provided by the radiation pressure.
Refs.~\cite{PRL02,Peng03,epl} considered the steady state of
different systems of driven cavities: Ref.~\cite{PRL02} focused on
two mirrors of a ring cavity and considered the situation in the
frequency domain; Ref.~\cite{Peng03} assumed to drive two
independent linear cavities with two-mode squeezed light and
stationary mechanical entanglement is achieved by transferring the
entanglement of the two driving beams to the end-mirrors of the two
cavities.

We have already partially compared the present scheme with that of
Ref.~\cite{epl}, to which is strongly related. In fact, the
double-cavity scheme of Ref.~\cite{epl} coincides with the single
Fabry-Perot cavity scheme considered here when the ``folded'' cavity
of Ref.~\cite{epl} is not driven. The additional folded cavity
couples to the center-of-mass of the two vibrational modes and if it
is appropriately driven by squeezed light, it is able to transfer
this squeezing to the center-of-mass. This has the advantage of
increasing the entanglement and making it more robust against
temperature (see Eq.~(\ref{logneg2})), but this is obtained at the
price of a more involved apparatus, requiring the preparation of an
additional ring cavity and the use of nonclassical driving.
Moreover, Ref.~\cite{epl} evaluated the stationary state of the two
mechanical modes approximately, by considering the resonant case
$\Delta=\Omega$ and solving the dynamics of the system only in the
limit when $\Omega$ is much larger than the other parameters, $G$,
$\kappa$, so that fast terms rotating at frequency $\Omega$ can be
neglected in the equations of motion. In this limit, Ref.~\cite{epl}
finds that the steady state of the two mirrors is entangled only if
the input field is squeezed, while is never entangled for a
classical coherent input. Here we determine the steady state of the
system \emph{exactly} in the Markovian limit of weak mechanical
damping and we find that when fast terms rotating at frequency
$\Omega$ cannot be neglected, \emph{one can entangle the mirror even
using classical driving}. As expected, it is possible to check that
the present exact solution reproduces the results of Ref.~\cite{epl}
in the same limiting conditions (no input squeezing, large
mechanical frequency, and no folded cavity). In fact, if we consider
$\Delta=\Omega \gg G,\kappa,\gamma $ in
Eqs.~(\ref{bip})-(\ref{diq}), one gets
\begin{eqnarray}\label{rwalimit}
&& b_p = b_q \simeq 0, \\
&& d_p= d_q = \frac{\gamma \left(G^2+2\gamma \kappa +4
\kappa^2\right)}{\left(\gamma+2\kappa\right)\left(G^2+2\gamma\kappa\right)},
\end{eqnarray}
so that
\begin{equation}\label{rwalimit2}
\langle q_{r}^2\rangle_{st}=\langle p_{r}^2\rangle_{st}=\frac{1}{2}+
\bar{n}\frac{\gamma \left(G^2+2\gamma \kappa +4
\kappa^2\right)}{\left(\gamma+2\kappa\right)\left(G^2+2\gamma\kappa\right)}
\geq \frac{1}{2},
\end{equation}
coinciding with Eq.~(23) of Ref.~\cite{epl} in the case of no input
squeezing, and implying absence of mechanical entanglement.
Therefore we see that the ``resonance'' condition $\Delta =\Omega$
is very close to the optimal condition for generating mechanical
entanglement, and that if one leaves the regime of vary large
mechanical frequencies $\Omega \gg G,\kappa,\gamma $, one can
achieve stationary mechanical entanglement even without input
squeezing. In fact, the parameter regime considered in
Figs.~\ref{entsi}-\ref{entno} corresponds to $\Omega \simeq G \gg
\kappa \simeq \gamma$.

Another recent proposal employing radiation pressure effects for
entangling two vibrating micro-mirrors is Ref.~\cite{entswap}, where
the radiation pressure of an intense laser field first generates
optomechanical entanglement between a mirror vibrational mode and an
optical sideband. Such an entanglement is then swapped to two
separated micro-mechanical oscillators via homodyne measurements on
the optical modes, representing Bell measurements in this continuous
variable setting. In this latter proposal, macroscopic mechanical
entanglement is generated when the homodyne measurement is performed
and it is therefore a transient phenomenon, with a lifetime which is
severely limited by the mirror thermal reservoir \cite{entswap}. In
the present scheme, on the contrary, mechanical entanglement has an
infinite lifetime because it is generated at the \emph{steady
state}, and therefore its experimental detection becomes much
easier.

We also notice that the system studied here is similar to the one
considered in Ref.~\cite{wien}, where a Fabry-Perot cavity with only
one vibrating mirror is considered. In Ref.~\cite{wien} a quantum
Langevin treatment analogous to the one adopted here is used to
quantify the amount of bipartite entanglement between the
vibrational mode of the mirror and the intracavity field at the
steady state of the system.

We finally discuss the experimental detection of the generated
mechanical entanglement. The measurement of $E_{\mathcal{N}}$ at the
steady state is quite involved because one has to measure all the
ten independent entries of the steady state correlation matrix $V$.
This has been recently experimentally realized (see
Ref.~\cite{laurat} and references therein) for the case of two
entangled optical modes at the output of a parametric oscillator.
Instead, one does not have direct access to the vibrational modes
and therefore it is not clear how to measure them. However
Ref.~\cite{wien} showed that, apart from additional detection shot
noise, the motional state of the mirror can be read from the output
of an adjacent Fabry-Perot cavity, formed by the mirror to be
detected and another ``fixed'' (i.e. with large mass) mirror. In
fact, it is possible to adjust the parameters of this second cavity
so that both position and momentum of the mirror can be
experimentally determined by homodyning the cavity output light
\cite{wien}. In particular, if the readout cavity is driven by a
much weaker laser so that its back-action on the mechanical mode can
be neglected, its detuning $\Delta_2$ is chosen to be equal to the
mechanical frequency $\Omega $, and its bandwidth $\kappa_2$ is
large enough so that the cavity mode adiabatically follows the
mirror dynamics, the output of the readout cavity $a_2^{out}$ is
given by
\begin{equation}
\label{output} a_2^{out}= i \frac{G_2 }{\sqrt{\kappa_2}}  b +
a_{2}^{in},
\end{equation}
where $b$ is the annihilation operator of the vibrational mode,
$G_2$ is the effective optomechanical coupling of the readout cavity
(see Eq.~(\ref{optoc})), and $a_2^{in}$ is the input noise entering
the readout cavity. Therefore using a readout cavity for each
mirror, changing the phases of the two local oscillators and
measuring the correlations between the two readout cavity output one
can then detect all the entries of the correlation matrix $V$ and
from them numerically extract the logarithmic negativity
$E_{\mathcal{N}}$ by means of Eqs.~(\ref{logneg}) and
(\ref{Sympl_eigenv}).

\section{Conclusions}

We have considered a system formed by a linear cavity with two
vibrating mirrors, driven by an intense classical light field. The
two mirror vibrational modes interact thanks to the radiation
pressure of the light bouncing between them. We have determined the
steady state of the system and we have seen that, in the case of
identical mechanical oscillators, the two vibrational modes become
entangled if the cavity detuning is close to the mechanical
frequency. The resulting mechanical entanglement is however quite
fragile with respect to temperature and this suggests that, in order
to generate macroscopic mechanical entanglement which is more robust
with respect to thermal effects, it is convenient to drive the
cavity with nonclassical light (see e.g., \cite{epl}.)

\section{Acknowledgments}

This work has been partly supported by the European Commission
through the Integrated Project Qubit Applications (QAP) funded by
the IST directorate, Contract No 015848 and by MIUR through PRIN-
2005 ``Generation, manipulation and detection of entangled light for
quantum communications''.


\Bibliography{<num>}

\bibitem{Nielsen}
M. A. Nielsen and I. L. Chuang, \textit{Quantum Computation and
Quantum Information} Cambridge University Press, Cambridge, 2000).

\bibitem{juuls01}
B. Julsgaard \textit{et al}., Nature (London) \textbf{413}, 400
(2001).

\bibitem{Berkley}
A. J. Berkley \textit{et al}., Science \textbf{300}, 1548 (2003).

\bibitem{LaHaye}M.D. LaHaye et al., Science \textbf{304},
74 (2004).

\bibitem{marsh}W. Marshall et al., Phys. Rev. Lett. \textbf{91}
130401 (2003).

\bibitem{Armour03}
A. D. Armour \textit{et al.}, Phys. Rev. Lett. \textbf{88}, 148301
(2002).

\bibitem{eisert04}
J. Eisert \textit{et al.}, Phys. Rev. Lett. \textbf{93}, 190402
(2004).

\bibitem{PRL02}
S. Mancini, V. Giovannetti, D. Vitali and P. Tombesi, Phys. Rev.
Lett. \textbf{88}, 120401 (2002).

\bibitem{Peng03}
J. Zhang \textit{et al.}, Phys. Rev. A \textbf{68}, 013808(2003).

\bibitem{dorsel}A. Dorsel, J. D. McCullen, P. Meystre, E. Vignes, and H. Walther,
Phys. Rev. Lett. \textbf{51}, 1550 (1983); A. Gozzini, F. Maccarone,
F. Mango, I. Longo, and S. Barbarino, J. Opt. Soc. Am. B \textbf{2},
1841 (1985).

\bibitem{bouwm}D. Kleckner \textit{et al.}, Phys. Rev. Lett. \textbf{96}, 173901
(2006).

\bibitem{arci1}O. Arcizet, P.-F. Cohadon, T. Briant, M. Pinard, A.
Heidmann, J.-M. Mackowski, C. Michel, L. Pinard, O. Francais and L.
Rousseau, Phys. Rev. Lett. \textbf{97} 133601 (2006).

\bibitem{gigan}S. Gigan, H. R. B\"ohm, M. Paternostro, F. Blaser, G. Langer, J.
B. Hertzberg, K. Schwab, D. B\"auerle, M. Aspelmeyer, A. Zeilinger,
Nature (London) \textbf{444}, 67 (2006).

\bibitem{arci2}O. Arcizet, P.-F. Cohadon, T. Briant, M. Pinard, and A.
Heidmann, Nature (London) \textbf{444}, 71 (2006).

\bibitem{wien}D. Vitali, S. Gigan, A. Ferreira, H. R. B\"ohm, P. Tombesi,
A. Guerreiro, V. Vedral, A. Zeilinger, and M. Aspelmeyer, Phys. Rev.
Lett. \textbf{98} 030405 (2007); M. Paternostro, D. Vitali, S.
Gigan, M. S. Kim, C. Brukner, J. Eisert, and M. Aspelmeyer, e-print
quant-ph/0609210.

\bibitem{kippen}T.J. Kippenberg, H. Rokhsari, T. Carmon, A. Scherer and K.J.
Vahala, Phys. Rev. Lett. \textbf{95} 033901 (2005); A. Schliesser,
P. Del' Haye, N. Nooshi, K.J. Vahala, T.J. Kippenberg, Phys. Rev.
Lett. \textbf{97} 243905 (2006).

\bibitem{epl}
M. Pinard, A. Dantan, D. Vitali, O. Arcizet, T. Briant and A.
Heidmann, Europhys. Lett. \textbf{72}, 747 (2005).

\bibitem{entswap}
S. Pirandola, D. Vitali, P. Tombesi, S. Lloyd, Phys. Rev. Lett.
\textbf{97}, 150403 (2006).

\bibitem{Pinard}
M. Pinard, Y. Hadjar, A. Heidmann, Eur. Phys. J. D \textbf{7}, 107
(1999).

\bibitem{law}
C. K. Law, Phys. Rev. A \textbf{51}, 2537 (1995).

\bibitem{gard}C. W. Gardiner and P. Zoller, \textit{Quantum Noise},
(Springer, Berlin, 2000).

\bibitem{Landau}L. Landau, E. Lifshitz, \textit{Statistical Physics} (Pergamon, New York, 1958).

\bibitem{GIOV01} V. Giovannetti, D. Vitali, Phys. Rev. A {\bf 63}, 023812 (2001).

\bibitem{simon}R. Simon, Phys. Rev. Lett. \textbf{84}, 2726 (2000).

\bibitem{noi}V. Giovannetti, S. Mancini, D. Vitali and P. Tombesi, Phys. Rev. A
\textbf{67}, 022320 (2003).

\bibitem{grad}I. S. Gradshteyn and I. M. Ryzhik, \textit{Table of Integrals, Series and Products}, Academic Press, Orlando, 1980, pag. 1119.

\bibitem{benguria} R. Benguria, and M. Kac, Phys. Rev. Lett, \textbf{46}, 1 (1981).

\bibitem{werner}G. Vidal and R. F. Werner, Phys. Rev. A \textbf{65}, 032314 (2002).

\bibitem{Salerno1}G. Adesso, A. Serafini, and F. Illuminati, Phys. Rev. A \textbf{70},
022318 (2004).

\bibitem{laurat}J. Laurat, G. Keller, J. A. Oliveira-Huguenin, C. Fabre, T. Coudreau, A. Serafini, G. Adesso,
and F. Illuminati, J. Opt. B: Quantum Semiclass. Opt. \textbf{7},
S577 (2005).



\endbib
\end{document}